\documentclass[12pt,amsmath,amssymb]{iopart}
\usepackage{latexsym}
\usepackage{graphics}
\usepackage{bm}
\usepackage{graphicx}

\begin{document}

\title{Static Black Hole dressed with a massive Scalar field}

\author{P.I.Kuriakose and V.C.Kuriakose}

\address{Department of physics, Cochin
university of Science and Technology, Kochi- 682022, India.}
\ead{bml@cusat.ac.in, vck@cusat.ac.in}

\begin{abstract}
A four-dimensional static black hole solution of Einstein equation
conformally coupled to a massive and self interacting scalar field
is obtained. A nontrivial scalar solution proposes a weak scalar
hair. A dressed black hole shows a trace of scalar charge in the
metric signalling the presence of scalar hair. A number of metrics
with regular horizons and temperatures are also proposed.

\end{abstract}

\maketitle

\section{Introduction}
No hair conjecture \cite{whee} demands the non existence of any
information other than mass, charge and angular momentum of a black
hole. There are  reports for and against the existence of scalar
hair. In order to prove the no - hair conjecture, no - hair theorems
had been established  by coupling the classical fields with the
Einstein gravity \cite{is}. It had been shown that the scalar field
would be trivial if one demands a regular horizon at a finite
distance from the center of  the  black  hole and also that
stationary black hole solutions are hairless in a variety of cases,
coupling different classical fields to gravity \cite{Xan,ch,be,te}.
Another calculation  \cite{Saa} showed that the static spherically
symmetric exterior solutions of the gravitational field equations in
a wide class of scalar tensor theories would essentially reduce to
the well known Schwarzschild solutions once the event horizon had
hidden the singularity. The above work was extended to charged black
holes and consolidated the non-existence of hair \cite{Sen}.

 As a weak interpretation of scalar hair, nontrivial
scalar field solution in terms of conserved charges was mooted
\cite{wein}. Using this ideology, scalar hair was reported in
asymptotically anti-de Sitter space time and asymptotically flat
space time and it is widely believed that black holes with scalar
hair generally exist when the scalar potential has negative region
\cite{tormae,nuca,Pik,bech}. There is no regular black hole solution
when the scalar field is massless or has a `` convex '' potential.
It is widely reported that there are no static asymptotically flat
and asymptotically AdS black holes with spherical scalar hair if the
scalar field theory, when coupled to gravity, satisfies the Positive
Energy Theorem \cite{tho}. A charged de Sitter black hole in the
Einstein-Maxwell-Scalar-$\wedge$ system possesses only unstable
solutions \cite{mak}. But an unexpected development of scalar hair
in AdS black hole with minimal \cite{tormae} as well as nonminimal
\cite{wis} coupling of scalar field, demanded a heuristic study of
scalar hair \cite{sud}.

As a strong interpretation, black hole has hair if there is a need
to specify quantities other than the conserved charges defined at
asymptotic infinity in order to characterize completely a stationary
black hole solution \cite{wein,biz}. Efforts were done to reveal
strong hair \cite{bocha,bek} and they came up with a scalar solution
conformally coupled to Einstein's gravity through a metric for
extremal case. Eventhough innocuous, the solution has a divergence
at the horizon. It is given as,
\begin{eqnarray}
\label{1}  \Phi=\frac{{-r_{0}}\alpha^{-1/2}}{r-r_{0}},
\end{eqnarray}
with $\alpha=\frac{8\pi G}{6}$. Eq. (1) is characterized by the ADM
mass, $\frac{r_{0}}{2}$ and scalar charge, $Q=4 \pi
r_{0}\alpha^{-1/2}$. But in Eq. (1), $\Phi$ blows up at the horizon,
which is against the principle that $\Phi$  shall be finite
everywhere\cite{suda}. In another attempt, a four dimensional
solution of the Einstein equation with a positive cosmological
constant coupled to a massless self interacting conformal scalar
field was put forwarded \cite{Zan}. The scalar solution in that case
is,

\begin{eqnarray}
\Phi(r)=\sqrt{3/4\pi}\frac{\sqrt{G}M}{r-GM}.
\end{eqnarray}
But Eq. (2) does not give any information other than mass of black
hole and the cosmological constant. Hence strong interpretation of
scalar hair is not guaranteed. The motivation of the present work is
to know whether a strong interpretation of the scalar hair can be
possible in a static (3+1) black hole, which requires a nontrivial
solution of scalar field in the vicinity of black hole with regular
horizon.

In this paper, we report a  nontrivial black hole solution of a
massive but self interacting scalar field showing no divergence at
the horizon and asymptotically falling to the vacuum value. The
proposed metric shows trace of scalar charge.

Whether a nontrivial scalar solution and a metric with a horizon are
mutually compatible has been the objective of scalar hair
investigation. Many contend that only when the solution is trivial
that a metric with a horizon is established and for a nontrivial
solution, the singularity will become naked. The criterion of scalar
hair is the coexistence of nontrivial solution and a proper metric (
having horizon and temperature )that holds the trace of scalar
field.

The scheme of the paper is as follows. In Sec. II, nontrivial scalar
hair solution is obtained. In Sec. III, the metric of the hairy
black hole is obtained. Sec. IV discusses thermodynamics of black
hole.  The conclusion is given in section V.

\section{Scalar hair Solution with a conformal coupling}

A nontrivial radial solution of a scalar field, whose source is a
scalar double well potential, in the vicinity of the static (3+1)
black hole will be discussed in this section. We will restrict our
consideration to the conformally coupled case. Consider the action,
\begin{eqnarray}
\label{3}I=\int d^4x\sqrt{-g}[\frac{R}{2k}-\frac 12g^{\mu \nu
}\nabla _\mu \Phi \nabla _\nu \Phi \\\nonumber-\frac 12\xi R\Phi
^2-\frac12 V(\Phi)],
\end{eqnarray}
where $\Phi $ is a massive, self interacting and conformally coupled
scalar field. For a (3+1) case, $ \xi=\frac16$. A double well
potential given by the equation,
\begin{equation}
 \label{4}V(\Phi)=-\frac12\mu^2(\Phi-\Phi_{0})^2+\frac14
 \delta^2(\Phi-\Phi_{0})^4+\frac14\frac{\mu^4}{\delta^2},
\end{equation}
has been introduced in the action given by Eq. (3).
\begin{figure}
\centering
\includegraphics[width=8cm]{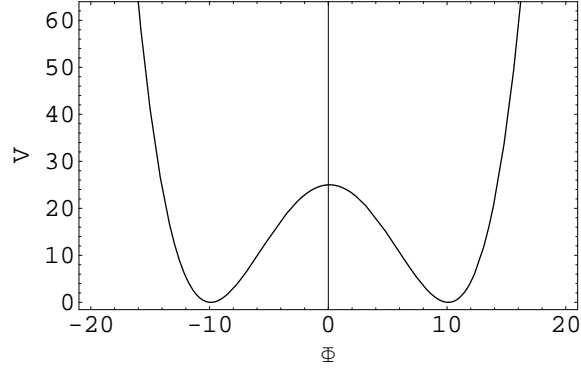}
\vspace*{8pt}\caption{ Variation of scalar potential against scalar
field $\Phi$, with, $\mu=1, \delta=0.1$  and $\Phi_{0}=0.1$. }
\end{figure}
The potential well is shown in Fig. (1). In Fig. (1), $V$ has global
minima at $\Phi=\pm\frac{\mu}{\delta}$ and a local maximum at
$\Phi=\Box\Phi_0$. The scalar field equation is given by,
\begin{equation}
 \label{5}\Box\Phi -\xi R \Phi +\frac12\mu^2(\Phi-\Phi_0) -\frac
12\delta ^2(\Phi-\Phi_{0})^3=0,
\end{equation}
where $\Box=g^{\mu\nu}\nabla_{\mu}\nabla_{\nu}$ is the
Laplace-Beltrami operator and $R$ represents the Ricci scalar. The
stress energy tensor of scalar field under gravity can be shown by
the relation,
\begin{equation}
\label{6}
\begin{array}{c}
T_{\mu \nu }=\nabla _\mu \Phi \nabla _\nu \Phi -\frac 12g_{\mu \nu
}g^{\alpha \beta }\nabla _\alpha \Phi \nabla _\beta \Phi +\\\\\frac
16[g_{\mu \nu }\Box
-\nabla _\mu \nabla _\nu +G_{\mu \nu }]\Phi^2 \\
\\
+\frac14g_{\mu \nu }\mu^2(\Phi-\Phi_0)^2-\frac 18g_{\mu \nu }\delta
^2(\Phi-\Phi_{0})^4-\frac18g_{\mu\nu}\frac{\mu^4}{\delta^2},
\end{array}
\end{equation}
with,
\begin{equation}
\label{7}
\begin{array}{c}
 \Box\Phi ^2=2\Phi  \Phi +2\nabla _\mu \Phi \nabla _\nu \Phi, \\
\\
\nabla _\mu \nabla _\nu \Phi ^2=2\Phi \nabla _\mu \nabla _\nu \Phi
+2\nabla _\mu \Phi \nabla _\nu \Phi.
\end{array}
\end{equation}
Here $\nabla_{\mu}$ represents a co-variant derivative.

For a static and spherically symmetric space time, the $t-t$
component of scalar stress energy tensor is given as,
\begin{equation}
\label{8}
\begin{array}{c}
T_0^0=-\frac 12g^{11}(\nabla _1\Phi )^2+\frac 16G_0^0\Phi^2
\\\\+\frac16\mu^2(\Phi-\Phi_0)^2-\frac1{12}\delta^2(\Phi-\Phi_{0})^4
-\frac1{12}\frac{\mu^4}{\delta^2},
\end{array}
\end{equation}
and the $r-r$ component of stress energy tensor is given as,
\begin{equation}
\label{9}
\begin{array}{c}
T_1^1=\frac 16g^{11}(\nabla _1\Phi )^2-\frac 13g^{11}\Phi \nabla
_1^2\Phi +\frac 13 (\nabla _1\Phi )^2 +\frac 16G_1^1\Phi^2\\
\\+\frac16\mu^2(\Phi-\Phi_0)^2-
\frac1{12}\delta^2 (\Phi-\Phi_{0})^4
-\frac1{12}\frac{\mu^4}{\delta^2}.
\end{array}
\end{equation}
The metric of a static (3+1) black hole may be given as,
\begin{equation}
\label{10}
ds^2=e^{2\nu}dt^2-e^{2\lambda}dr^2-r^2d\theta^2-r^2\sin^2\theta
d\varphi^2.
\end{equation}
In the above metric, $\lambda$ is a function of $r$ only and
$\nu=\lambda(r)+f(t)$. Here, $f(t)$ is an arbitrary function of $t$.
There is no loss of generality in setting $f(t)=0$, since it can be
absorbed in the definition of $t$, i.e., by replacing $e^{f(t)}dt$
by $dt$. With this redefinition of the time coordinate,
$\nu=-\lambda$. Then,
\begin{equation}
\label{11} G_0^0=G_1^1=\frac1{r^2}\frac d{dr}[r(1-e^{2\nu})].
\end{equation}
We know that black hole is a thermodynamical system with a
temperature and its thermal radiation is characterized by
renormalized stress-energy tensor  \cite{page}. The gravitational
effect of heat radiation, characterized by its gravitationally
induced renormalized stress-energy tensor, is constructed on real
Euclidean section of the black hole geometry with its (Euclidean)
time coordinate identified with period $\beta_{0}=8\pi M$ ( for
S.B.H), so as to eliminate singularity at the horizon. Applying Eq.
(11) in Einstein's equation $G^{\mu}_{\nu}- \kappa T^{\mu}_{\nu}=0$,
we get,
\begin{equation}
\label{12} T_0^0-T_1^1=0.
\end{equation}
The concept of scalar hair is applicable only in a static black
hole. From Eq. (12), we get,
\begin{equation}
\label{13}-2g^{11}(\nabla _1\Phi )^2-(\nabla _1\Phi )^2+ g^{11}\Phi
\nabla _1^2\Phi =0.
\end{equation}
Eq. (13) is a co-variant differential equation. In the above scalar
field equation, properties such as  mass and self interaction terms
of scalar field do not come explicitly. In Eq. (13), $\nabla
_1^2\Phi$ can be written in the ordinary derivative as,
\begin{equation}
\label{14} \nabla _1^2\Phi=\partial _1^2\Phi -\Gamma _{11}^i\partial
_i\Phi ,
\end{equation}
where $\Gamma$ is the usual Christoffel symbol and $i$ runs from $0
\rightarrow 3$. In the above case, all the Christoffel symbols
except $\Gamma _{11}^1$ are zeroes and $\Gamma _{11}^1=
{\lambda}^{'}=- {\nu}^{'}$. Now, Eq. (13) gets modified as,
\begin{equation}
\label{15}-2g^{11}(\partial _1\Phi )^2-(\partial _1\Phi )^2+
g^{11}\Phi [\partial _1^2\Phi +\partial_1\nu \partial _1\Phi ]=0.
\end{equation}
In quest of scalar hair, the general principle is to find a
nontrivial solution which is compatible with a proper metric. So we
propose a solution of the form,

\begin{equation}
\label{16}\Phi(r)=a sech[arccosh[r]]+\Phi_{0},
\end{equation}
where $\Phi_{0}$ is the asymptotic value of scalar field, $'a'$ is a
constant which may be derived from mass of black hole and scalar
field. It can be shown that $\Phi(r)=a
sech[arccosh[r]]+\Phi_{0}=\frac{a}{r}+\Phi_0$. Nontrivial solution
of scalar field have been proposed by many people earlier but never
extended it to the concept of scalar hair \cite{Zan,Cri}. By
substituting Eq. (16) in Eq. (15), we get,
\begin{equation}
\label{17}\Phi_{0}g^{11}\partial _1^2\Phi-(\partial _1\Phi )^2+
g^{11} \Phi
\partial_1\nu\partial _1\Phi =0.
\end{equation}
Our primary aim is to evolve a metric with function $e^{2\nu}$,
which is equal to zero at the horizon and equal to $1$ $(
i.e.,\nu=0)$ in the asymptotic limit. In Eq. (17), $\Phi_{0}$ is a
very small field constant, $g^{11}$ is zero at the horizon and
$\partial_1^2\Phi$ is zero in the asymptotic limit. So the term,
 $\Phi_{0}g^{11} \partial_1^2\Phi$  is extremely small through out the space and hence may be neglected.
 Eq. (17)  can be
 rewritten as,
\begin{equation}
\label{18}-\partial _1\Phi + g^{11} \Phi
\partial_1\nu =0.
\end{equation}
 From Eq. (18), we can determine
 the metric function which is given in the Sec.111.
   The profile of Eq. (16) shows that the field has a finite value
  $\Phi_{h}$ at the horizon
  and then falls to the asymptotic value $\Phi_{0}$. The variation
of $\Phi$ against $r$ is shown in Fig. (2).

\begin{figure}
\centering
\includegraphics[width=8cm]{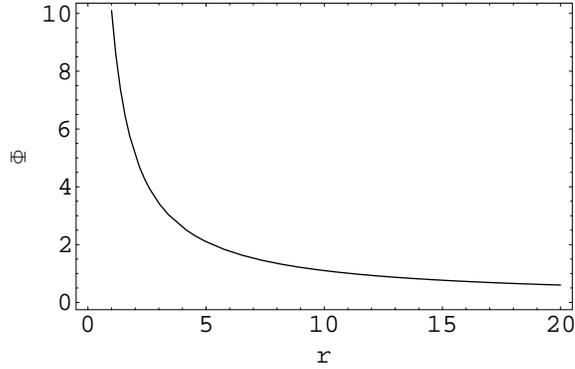}
\vspace*{8pt}\caption{ Variation of scalar field against $r$ with
$a=1$ and $\Phi_{0}=0.1$.}
\end{figure}

\section{Metric of the (3+1) black hole}

The form of metric, which is compatible with the scalar solution,
will be determined. Since $g^{11}=-e^{2\nu}$, Eq. (18)becomes,
\begin{equation}
\label{19}\partial _1\Phi +e^{2\nu}\Phi
\partial_1 \nu =0.
\end{equation}
By introducing a transformation of the type \cite{kis},
\begin{equation}
\label{20}\nu=\frac12\log(1+f),
\end{equation}
we get from Eq. (19),
\begin{equation}
\label{21}\frac{d\Phi}{\Phi}=-\frac12 df,
\end{equation}
where $f$ is a radial function. Integrating Eq. (21),
\begin{equation}
\label{22}\log\Phi=-\frac{f}2+C_{0}.
\end{equation}
In the asymptotic limit $\nu=0$ and hence $f=0$.  Putting the
asymptotic value of $\Phi$ as $\Phi_0$, we can obtain,
\begin{equation}
\label{23}\log(\frac{\Phi}{\Phi_{0}})=-\frac{f}{2}.
\end{equation}
Thus from Eqs. (20) and (23) we find,
\begin{equation}
\label{24}e^{2\nu}=1-2\log(\frac{\Phi}{\Phi_{0}}).
\end{equation}
Eq. (24) represents the metric function which is compatible with the
scalar solution $\Phi(r)=a
sech[arccosh[r]]+\Phi_{0}=\frac{a}{r}+\Phi_0$. In the asymptotic
limit the metric function,
\begin{equation}
\label{25}e^{2\nu}=1,
\end{equation}
coincides with those of SBH and RN black holes. Denoting the field
at the horizon as $\Phi_h=\Phi_0 e^{1/2}$, the radius of the horizon
can be obtained as:

\begin{equation}
\label{26}r_{h}=\frac{a}{{\Phi_{0}}(e^{1/2}-1)}.
\end{equation}
In Eq. (24), $\log(\frac{\Phi}{\Phi_{0}})$ can be expanded as a
series if $\frac{\Phi}{\Phi_{0}}\leq 2$ and it is true in the
present case. Therefore,
\begin{equation}
\label{27}\log(\frac{\Phi}{\Phi_{0}})=(\frac{\Phi}{\Phi_{0}}-1)
-\frac12(\frac{\Phi}{\Phi_{0}}-1)^2+\frac13(\frac{\Phi}{\Phi_{0}}-1)^3....
\end{equation}
The metric function then may be written as a series as,
\begin{equation}
\label{28}e^{2\nu}=1-2(\frac{\Phi}{\Phi_{0}}-1)
+(\frac{\Phi}{\Phi_{0}}-1)^2-\frac23(\frac{\Phi}{\Phi_{0}}-1)^3....
\end{equation}
In Eq. (28), let $(\frac{\Phi}{\Phi_{0}}-1)
=\frac{a}{\Phi_{0}r}=\frac{b}{r}$. The metric function gets modified
as,
\begin{equation}
\label{29}e^{2\nu}=1-\frac{2b}{r}+\frac{b^2}{r^2}-\frac23
\frac{b^3}{r^3}...
\end{equation}
The above mentioned metric is in unison with a recent
work\cite{Eva}.

\subsection{Study of metric}
 Eq. (28) may be written in a
concise form as,
\begin{equation}
\label{30}e^{2\nu}=1-2[\sum_{n=1}^{n=\infty}(-1)^{n+1}\frac{1}{n}(\frac{\Phi}{\Phi_{0}}-1)^n].
\end{equation}
When $n=1$, we have the Schwarzschild like black hole:
\begin{equation}
\label{31}e^{2\nu}=1-2(\frac{\Phi}{\Phi_{0}}-1)=1-\frac{2b}{r}.
\end{equation}
In this case, $\frac{\Phi_{h}}{\Phi_{0}}=\frac32$. The radius of
horizon is $r_{h}=2\frac{a}{\Phi_0}=2b$.

 When $n=2$, we have the extremal case similar to
extremal RN black hole. The metric function  is reduced to,
\begin{equation}
\label{32}
\begin{array}{c}
e^{2\nu}=1-2(\frac{\Phi}{\Phi_{0}}-1)+(\frac{\Phi}{\Phi_{0}}-1)^2=
[1-(\frac{\Phi}{\Phi_{0}}-1)]^2 \\\\
=(1-\frac{b}{r})^2.
\end{array}
\end{equation}
In the extremal case, $\frac{\Phi_{h}}{\Phi_{0}}=2$, which gives the
maximum value of $\frac{\Phi_{h}}{\Phi_{0}}$. The radius of horizon
is $r_{h}=\frac{a}{\Phi_0}=b$.

When $n=3$, $\frac{\Phi_{h}}{\Phi_{0}}=\frac{13}{8}$ and
$r_{h}=\frac53\frac{a}{\Phi_0}=\frac53 b$. When $n=4$,
$\frac{\Phi_{h}}{\Phi_{0}}=\frac{17}{10}$ and
$r_{h}=\frac23\frac{a}{\Phi_0}=\frac23 b $. It is obtained that the
horizon's radius increases and decreases with diminishing magnitude
as $n$ increases. We can thus introduce more types of black holes by
putting $n=5,6 ... $. But as the series progress, the series very
quickly diminishes. The variation of scalar field $\Phi$ against $r$
for different black holes are shown in Fig. (3). The thick line
represents the case with $n=1$. The normal line graph represents the
case with $n=2$. The dashed line represents the case with $n=3$ and
the dotted line represents the case with $n=4$.
\begin{figure}
\centering
\includegraphics[width=8cm]{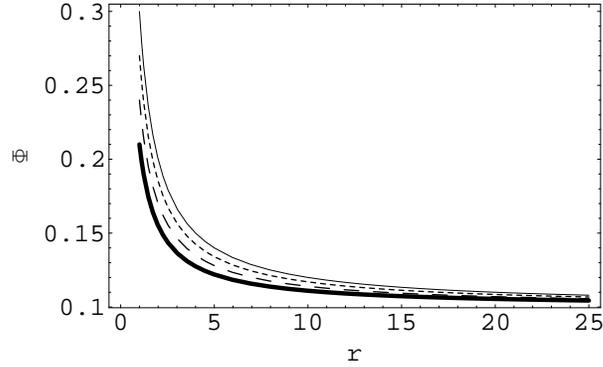}
\vspace*{8pt}\caption{ Variations of scalar field $\Phi$ against $r$
for different black holes, with, $\Phi_{0}=0.1$.}
\end{figure}

 We know that the Hawking's evaporation eventually finishes a black hole.
 To avoid this situation, a cavity was proposed to contain the black hole so that Hawking
radiation and quantum field exist in equilibrium inside the cavity
\cite{york}. The black hole with a cavity is called dressed black
hole and without it is called naked black hole.

The signature of scalar field in the metric is due to the asymptotic
value of scalar tensor $T^0_0$. The asymptotic value of $T^0_0$ is
given as,
\begin{equation}
\label{33}T^0_0=\frac16G^0_0(\infty)\Phi_0^2-\frac1{12}\frac{\mu^4}{\delta^2}.
\end{equation}
As $G^0_0$ vanishes in the asymptotic limit we get,
\begin{equation}
\label{34}T^0_0=-\frac1{12}\frac{\mu^4}{\delta^2}.
\end{equation}
The mass term corresponding to scalar field may be written as,
\begin{equation}
\label{35} \eta(r)=-\int_{r_h}^{r_0} 4 \pi r^2T^0_0dr.
\end{equation}
Substituting Eq. (34) in Eq. (35), we get,
\begin{equation}
\label{36}\eta=\frac{\pi}{9} \frac{\mu^4}{\delta^2}(r_0^3-r_h^3).
\end{equation}
The $\eta$ of Eq. (36) has its contribution in the making of
signature of scalar field in the metric.

\subsection{Mass of hairy black hole}
 Mass $(m_{rh})$ of a hairy black hole is related to the mass
 $(M_{rh})$
  of a non-hairy black hole
 \cite{Pik,sud} through the relation,
\begin{equation}
\label{37}m_{r_h}=M_{r_h}+4\pi r^2\int_{r_h}^rr[V(\Phi )-V(\Phi
_\infty )+\frac 12e^{2\nu} \Phi ^{^{\prime }2}]dr.
\end{equation}
In Eq. (37), $V(\Phi)=0$  at the horizon and
$V(\Phi_{\infty})=\frac14\frac{\mu^4}{\delta^2}$. As the distance
from the center of black hole increases, $m(r_h)$ increases, but
never blows up since it attains a steady value as $r$ increases. The
variation of $m(r_h)$ against $r$ is shown in the Fig. (4).

\begin{figure}
\centering
\includegraphics[width=8cm]{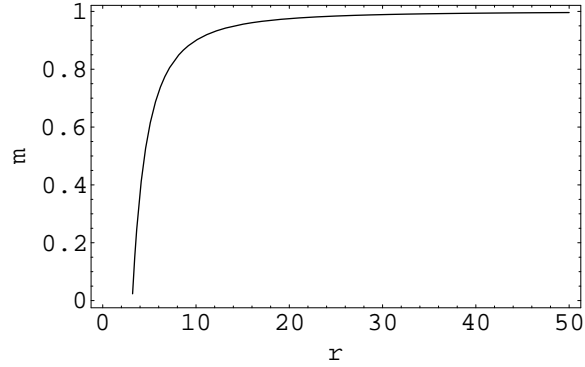}
\vspace*{8pt}\caption{ Variation of mass of hairy black hole against
$r$, with, $a=1$ and $\Phi_{0}=0.1$.}
\end{figure}
\subsection{Entropy } Surface gravity of the black hole is given
as,
\begin{equation}
\label{38}k=\frac{\partial _rg_{tt}}{2\sqrt{-g_{tt}g_{rr}}}\mid
_{r=r_h}.
\end{equation}
With $g_{tt}=e^{2\nu}=1-2 \log(\frac{\Phi}{\Phi_0})$ and
$\Phi(r)=\frac a{r}+\Phi_0$, we get,
\begin{equation}
\label{39}k=\frac a{a r_h+\Phi_0 (r_h)^2}.
\end{equation}
Substituting for $r_h=\frac a{\Phi_0(e^{1/2}-1)}$ in Eq. (39), we
find the temperature of black hole as,
\begin{equation}
\label{40}T_b=\frac {\Phi_0(e^{1/2}-1)^2}{2\pi a e^{1/2}}.
\end{equation}
The area of the horizon,
\begin{equation}
\label{41}
\begin{array}{c}
A_h=4\pi r_h^2=4\pi \frac{a^2}{\Phi_0^2(e^{1/2}-1)^2},
\end{array}
\end{equation}
exclusively depends on the parameter 'a'. Following the standard
procedure, we can determine the entropy of black hole in the
presence of scalar field.
\begin{equation}
\label{42}
\begin{array}{c}
dA_h= \frac4{e^{1/2}\Phi_0}dS.
\end{array}
\end{equation}
On integrating,
\begin{equation}
\label{43}
\begin{array}{c}
S=\Phi_0e^{1/2} \frac{A_h}4+c=\Phi_h \frac{A_h}4+c,
\end{array}
\end{equation}
where $c$ is a constant and $S$ the entropy of black hole in the
presence of scalar field. By applying the boundary condition that
when $\Phi_h=0$, $S=S_0$, the entropy of naked black hole. Then
$c=S_0$. But by Hawking's theory, $S_0=\frac{A_h}{4}$. So the total
entropy is written as,
\begin{equation}
\label{44} S=\Phi_h \frac{A_h}4+\frac{A_h}4=S_s+S_0.
\end{equation}
Eq. (44) clearly indicates that the scalar field contributes to the
entropy of black hole.

\section { Thermodynamics}

Hawking's discovery that a black hole radiates energy with a thermal
spectrum prompted the scientific community to believe that a black
hole can exist in thermal equilibrium with a heat bath composed of
quantum fields interacting with the black hole geometry. When the
Hawking radiation is fully thermal, the thermal pressure is
$\frac13\alpha T_{loc}^4$, where
$T_{loc}=\frac{T_{b}}{\sqrt{-g_{00}}}$. The stress-energy tensor of
the radiation is a function of black hole temperature $T_{b}$. The
hairy black hole is a thermodynamical system with a modified
temperature.

The effective potential of the test particles moving in static and
spherically symmetric background geometry is determined by the
Hamilton-Jacobi approach. By Hamilton-Jacobi equation,
\begin{equation}
\label{45}g^{k\lambda }\partial _kS\partial _\lambda S+\mu ^2=0,
\end{equation}
where $\mu$ is the mass of scalar field and,
\begin{equation}
\label{46}S(t,r,\phi )=-Et+S(r)+L\phi.
\end{equation}
$E$ and $L$ are the constant energy and angular momentum of the test
particle. Substituting Eq. (46) in Eq. (45) and simplifying we get,
\begin{equation}
\label{47}S(r)=\mp \int [E^2-(\frac{L^2}{r^2}+\mu ^2)f]^{1/2}\frac
1fdr,
\end{equation}
where $f=e^{2\nu}$. \subsection{Special cases}

(a). In Eq. (30), with $n=1$,
$e^{2\nu}=1-2(\frac{\Phi}{\Phi_{0}}-1)=1-\frac{2b}{r}$. This is SBH
like. As $r\rightarrow r_{h}$, $S(r)$ is modified as,
\begin{equation}
\label{48}S(r)=\mp\int\frac{r_{h} E
dr}{r-r_{h}}=\mp\beta\log(r-r_{h}).
\end{equation}
with $\beta=E r_{h}$. Assuming that the scalar field gets reflected
at the horizon, the scalar field in the neighborhood of the horizon
can be written as \cite{book},
\begin{equation}
\label{49}\Phi (r)=e^{-i\beta \log{(r-r_h)}}+R e^{i \beta
\log{(r-r_h)}},
\end{equation}
with $R$ as the coefficient of reflection. The coefficient of
reflection $R$ and the probability of reflection by horizon $P$ may
be given as,
\begin{equation}
 \label{50}R=e^{-2\pi\beta}; P=\mid R\mid ^2=e^{-4\pi \beta}.
\end{equation}
Using the thermodynamical relation,
\begin{equation}
\label{51}P=e^{-E/T_{b}},
\end{equation}
we get,
\begin{equation}
\label{52} E/ T_{b}=4\pi\beta,
\end{equation}
which gives the black hole temperature as,
\begin{equation}
\label{53}T_{b}=\frac1{4\pi  r_{h}}.
\end{equation}

This is the temperature of Schwarzschild like black hole.\\\\

(b). In Eq. (30), with  $n=2$,
$e^{2\nu}=[1-(\frac{\Phi}{\Phi_{0}}-1)]^2=[1-\frac{b}{r}]^2$. This
is extremal case in which no black hole temperature is observed.\\\\

(c). In Eq. (30), with $n=3$,
$e^{2\nu}=1-2(\frac{\Phi}{\Phi_{0}}-1)+(\frac{\Phi}{\Phi_{0}}-1)^2-
\frac23(\frac{\Phi}{\Phi_{0}}-1)^3=1-\frac{2b}{r}+\frac{b^2}{r^2}-\frac23
\frac{b^3}{r^3}$. The action $S(r)$ as $r\rightarrow r_{h}$ is given
as,
\begin{equation}
\label{54}S(r)=\mp\int \frac{E r_{h}^3 dr
}{(r-r_{h})(r_{c}-r_{h})(r_{h}-r_{0})},
\end{equation}
with, $\beta=\frac{E r_{h}^3 }{(r_{c}-r_{h})(r_{h}-r_{0})}$ and
$(r_{c}+r_{h}+r_{0})=-2b; (r_{h}r_{c}+r_{c}r_{0}+r_{b}r_{0})=-b^2;
r_{h}r_{c}r_{0}=-\frac23b^2$. Now,
\begin{equation}
\label{55}S(r)=\mp \beta \log(r-r_{h}).
\end{equation}
 By proceeding as in the
previous case, the temperature of the black hole can be shown to be,
\begin{equation}
\label{56}T_{b}=\frac{(r_{c}-r_{h})(r_{h}-r_{0})}{4\pi r_{h}^3} .
\end{equation}
Thus we see that the hairy black hole acts as a thermodynamical
system with proper horizon and temperature.
\section {conclusion}

The scalar hair of a black hole is a bone of contention. There is a
general belief that anything that is added to a black hole will not
induce any trace of it on the black hole except mass, angular
momentum and vector charge and hence `` no-hair conjecture''. There
is an argument that a regular horizon is possible only when the
scalar solution is trivial\cite{Saa,Sen}. So when the solution is
nontrivial, the event horizon will be a surface of singularity and
hence it will not represent a black hole. A black hole exists only
when the event horizon hides the naked singularity.

As a weak interpretation of scalar hair, a nontrivial solution of
scalar field in terms of the existing conserved quantities is enough
to show that there is hair \cite{tormae,nuca,Pik}. Whether a horizon
naturally occurs, even when the solution is nontrivial, will be the
primary objective of strong interpretation of scalar hair. Thus as a
strong interpretation, in the presence of a scalar field, a black
hole can have a signature different from mass, angular momentum and
vector charge.

We have shown that a nontrivial scalar black hole solution for a
massive self interacting conformal scalar field would be obtained in
the case of static (3+1) black hole. The metric proposes horizon and
temperature for the black hole. The horizon and surface temperature
ensure a true black hole. The metric element has a term other than
the existing conserved quantities such as mass, vector charge and
angular momentum. In our case, only a particular pair of scalar
field and metric are found to be mutually compatible.

In the proposed metric, only one parameter,i.e.,  $b$ has appeared.
This may invite some criticisms against the strong interpretation of
scalar hair. But in the standard extremal case, same parameter does
the job of mass and vector charge. Another argument against $b$ is
that it may have been originated from the mass of black hole itself.
But Eq. (37) clearly indicates the origin of the parameter $b$ is
from scalar field and hence we may conclude that scalar field can
depict its signature in the metric.
\subsection{Acknowledgments}
We would like to thank J. Zanelli for useful discussions. V.C.K
wishes to acknowledge UGC, New Delhi, for financial support through
Major Research project and  Associateship of IUCAA, Pune.

\end{document}